\documentstyle[12pt]{article}
\begin{document}

\thispagestyle{empty}
\begin{flushright}                              FIAN/TD/99-2\\
                                                hep-th/9904215\\
                                                April 1999
\vspace{2cm}
\end{flushright}

\begin{center}
{\Large S. S. Shahverdiyev$^{a}$, I. V. Tyutin$^{b}$, B. L.
Voronov$^{c}$ } \end{center}

\vspace{2cm}

{\large \bf{On Local Variational Differential Operators in Field Theory} }

\vspace{5ex}

{\large Department of Theoretical Physics, P.~N.~Lebedev Physical
Institute,

  Leninsky prospect 53, 117924 Moscow, Russia}
\vspace{10ex}

\centerline{{\Large\bf Abstract}}
\begin{quote}
{\normalsize
We propose and develop a new calculus for local variational differential operators.
The main difference of the new formalism with the canonical
differential calculus is that the image of higher order
operators on local functionals does not contain
indefinite quantities like $\delta(0)$.  We apply this formalism to BV
formulation of general gauge field theory and to its $Sp(2)$-symmetric
generalization.  Its relation to a quasiclassical expansion is also
discussed.}
\end{quote}

\vfill
\noindent
$^a$ E-mail address: shahv@td.lpi.ac.ru

\noindent
$^b$ E-mail address: tyutin@td.lpi.ac.ru

\noindent
$^c$ E-mail address: voronov@td.lpi.ac.ru

\newpage
\section{Introduction}

In the functional formulation of local quantum field theory,
it is rather common to encounter a problem that is known as
"the problem of $\delta(0)$". The essence of the problem can be
illustrated by the following simple
example.  Let $S(\varphi)$ be a local functional of a field $ \varphi(x)$,
say the action
$$
S(\varphi)=\int dx{\cal L}(\varphi(x),\partial_\mu\varphi(x),...),\quad
\partial_\mu={\partial\over\partial x_\mu}.
$$
Then the second order variational derivative $\delta^2S(\varphi)/
\delta\varphi(x_1)\delta\varphi(x_2)$ of this functional is a quasilocal
distribution (for the notion of a quasilocal distribution see subsection 2.3),
symbolically,
$\delta^2S(\varphi)/\delta\varphi(x_1)\delta\varphi(x_2)\sim\delta(x_1-x_2)$.
Therefore, the second order variational differential operators
of the type
$$
\Delta_2 =\int dx E_2(\varphi(x), \partial_\mu \varphi(x),...)
\frac{\delta}{\delta\varphi(x)}\frac{\delta}{\delta\varphi(x)}
$$
are not defined on local functionals: $\Delta_2S\sim\delta(0)$. These operators
are not defined on the products $S_1(\varphi)S_2(\varphi)\ldots$ of local
operators, and generally on the functions $f(S_1,S_2,\ldots)$ of the latter
ones, if we assume the standard Leibnitz rule $$
\Delta_2(S_1(\varphi)S_2(\varphi)\cdots{}) =
(\Delta_2S_1(\varphi))S_2(\varphi)\cdots{}+2\int dx E_2(...) \frac{\delta
S_1}{\delta\varphi(x)}\frac{\delta
S_2}{\delta\varphi(x)}\cdots{}+
$$
$$
S_1(\Delta_2S_2)\cdots{}+....
$$
The analogous assertions are valid for the higher order differential derivatives
\\$\delta^nS(\varphi)/\delta\varphi(x_1)\ldots\delta\varphi(x_n)$, $n>2$,
and therefore for the local higher order differential operators
$$
\Delta_n=\int dx E_n(\varphi(x),
\partial_\mu\varphi(x),...)(\delta/\delta\varphi(x))^n.
$$
However, the emergence of such operators is unavoidable in local field theory.
For instants, they arise when we consider the change of variables in
the functional integrals that define a theory, in particular, when we seek
an invariant measure for local general gauge field theory. Moreover, the
$\Delta_2$-type operator underlies the BV formulation of gauge theories
\cite{BV}. The $\delta(0)$-terms that arise in local theories must be
compensated by the corresponding "local measure" that is initially singular.
It can happen, as ``for example'' in Yang -- Mills or supersymmetric
theories, that the coefficient at $\delta(0)$ is equal to zero from the very
beginning because of specific algebraic reasons, then the indefiniteness
$0\times\infty$ is resolved in favor of the zero.

In a general case, the standard way of avoiding the problem of $\delta(0)$
consists in chattering something like this:"if using dimensional regularization
\cite{Dim}, the corresponding singularity
$\sim \delta(0)$  is equal to zero."
In other words, the formal rule``$\delta(0)=0$'' is adopted at intermediate
stages of the formulation.  The successful experience of operating with this
rule makes us to suggest the possibility of such a mathematically consistent
formalism, where $\delta(0)$ does not arise at all. In this paper we realize
such a possibility in the framework of a formalism that is proposed below.
In this formalism, the absence of $\delta(0)$ is a consequence of the
definitions. First of all, the proposed formalism is applied to the
BV formulation of general gauge field theory. It turns out that all the
contents of the theory remains unchanged.  The proposed formalism is
specific just for local field theory and has no direct analogs in the
theory of functions of finite number of variables.  This fact cautions once
more against possible attempts to justify some, basic sometimes, assertions
in local field theory by referring to the finite-dimensional analogs,
and even more, to replace the proofs by such references \cite{BT}.
On the other hand, as it will be seen below, the proposed formalism
furnishes the consistent quasiclassical (with respect to the Plank
constant $\hbar$) quantization.

The paper is organized as follows. In sec. 2, we define the appropriate
classes of functionals and operators, formulate the rules of the action of the
operators on the functionals and discuss their properties. In sec. 3, we
consider the special class of operators of the type $\exp\nabla$ and  the
special class of functionals of the type $\exp S$, as well as the relation of
the proposed formalism to the quasiclassical expansion. In sec.4, we consider
the local changes of variables.  In sec.5 and sec.6 respectively, we apply
the proposed formalism to the Lagrangian BV formulation of gauge theories and
to its $Sp(2)$-symmetric generalization.

\section{Functionals, operators $\nabla$}

\subsection{Fields}

We let $\Gamma=\{\Gamma^A(x)\}$ denote the initial fields and use the condensed
notations like $\Gamma^{A_k}=\Gamma^{A_k}(x_k)$,
$E_n^{(A)_n}=E_n^{A_1...A_n}(x_1,\ldots,x_n;\Gamma)$. The fields take its
values in the Berezin algebra that includes Bose and Fermi fields on equal
footing,their Grassmann parities are denoted by
$\varepsilon(\Gamma^A)=\varepsilon_A$. The derivatives with respect to the
fields are left,
$$
\frac{\delta}{\delta\Gamma^{A_k}(x_k)}=\delta_{A_k}, \quad
\varepsilon(\delta_{A_k})=\varepsilon_{{A_k}}
$$
In addition, we use the following abbreviations:
$$
(\delta_A)^n=\delta_{A_1}\ldots\delta_{A_n},\quad
\varepsilon((\delta_A)^n)=\sum_{k=1}^n\varepsilon_{A_k},
$$
$$
(A)_m=A_1\ldots A_m.
$$
The index, which appears twice in the same term, implies a summation
(integration).

\subsection{Functionals}

The essential starting point is the restriction on the class of the
functionals $\Phi$ of the fields $\Gamma$. First of all, it
includes all smooth local functionals $S_i(\Gamma), i=1, 2,...$, with the
certain Grassmann parity $\varepsilon(S_i)=\varepsilon(i)$, of the type
\begin{equation}\label{1} S_i(\Gamma)=\int dx {\cal
L}_i(x,\Gamma(x),\partial_\mu\Gamma(x),...), \end{equation}
where a local density $ \cal L_i$ is a smooth function of coordinates $x$,
fields $\Gamma(x)$ and their finite order derivatives.  The locality of the
functional $S(\Gamma)$ implies that its variational derivatives
$(\delta_A)^nS(\Gamma)$ $=$
$\delta/\delta\Gamma^{A_1}(x_1)\ldots\delta/\delta\Gamma^{A_k}(x_k) S(\Gamma)$
are quasilocal distributions, symbolically
 $$
 (\delta_A)^nS(\Gamma))\sim \delta(x_1-x_2)\cdots
 \delta(x_{n-1}-x_n).
$$
The functionals $\Phi$ of a general type are smooth functions of local
functionals:
\begin{equation}\label{2}
\Phi(\Gamma)=f(S_1, S_2,...)=f(S),
\end{equation}
usually with the certain Grassmann parity $\varepsilon(f)$; we will call them
the multilocal functionals. The derivatives with respect to $S_i$ are left,
$$
\frac{\partial}{\partial S_i}=\partial_i,
\quad\varepsilon(\partial_i)=\varepsilon(i).
$$

In the sequel, we use the following natural abbreviations in our notations:
$$
(S_i)^n=S_{i_1}\cdots S_{i_n},\quad
\varepsilon((S_i)^n)=\sum_{k=1}^n\varepsilon(i_k),
$$
$$ (\delta_AS_i)^n=\delta_{A_1}S_{i_1}\cdots\delta_{A_n}S_{i_n},\quad
\varepsilon((\delta_AS_i)^n)=\sum_{k=1}^n\varepsilon_{A_k}+
\sum_{k=1}^n\varepsilon(i_k),
$$
$$
(\delta_A\circ S_i)^n=(\delta_AS_i)^n(-1)^{\sum_{j=1}^{n-1}\varepsilon(i_j)
\sum_{k=j+1}^n\varepsilon_{A_k}},
$$
$$
f_,{}_{i_1\ldots i_n}=\partial_{i_1}\ldots\partial_{i_n}f.
$$
$$
(i)_n=i_1\ldots
i_n,\quad\overleftarrow{(i)_n}=i_n\ldots i_1,
$$
such that for instance
$$
f,_{(i)_n}=f,_{i_1\ldots i_n},\quad
f,_{\overleftarrow{(i)_n}}=f,_{i_n\ldots i_1},
$$

\subsection{Local variational differential operators $\nabla$}

On the functional space considered, we define the multilocal operators,
the epithet ``multilocal'' means that the image of the operator lies in the
same space of multilocal functionals.  We begin with the homogenous local
differential operators $\nabla_n$ of an arbitrary order $n$. Let us consider
the local differential expression \begin{equation}\label{3} \nabla_n=\int
dx_1\ldots dx_nE_n^{A_1...A_n}(x_1,\ldots,x_n;\Gamma)
\frac{\delta}{\delta\Gamma^{A_1}(x_1)}...
\frac{\delta}{\delta\Gamma^{A_n}(x_n)}=
 E_n^{(A)_n}(\delta_A)^n,
\end{equation}
with the certain Grassmann parity $\varepsilon(\nabla_n)$, and with the
coefficient functions
\begin{equation}\label{4}
E_n^{A_1...A_n}(x_1,...,x_n;\Gamma)=
\sum_{m_1...m_n}\int dxE_{n|m_1...m_n}^{A_1...A_n}
(x,\Gamma(x),\partial_\mu\Gamma(x),...)
\prod_{k=1}^n(\partial_\mu)^{m_k}\delta(x-x_k),
\end{equation}
where $E_{n|m_1...m_n}^{A_1...A_n}(x,\Gamma(x),\partial_\mu\Gamma(x),...)$
are smooth functions of their arguments. The coefficient functions \\
$E_n^{A_1\ldots A_n}(x_1,\ldots,x_n;\Gamma)$ $=$ $E_n^{(A)_n}$
are distributions with respect to the space arguments $x_1,\ldots,x_n$
with the support at the coinciding points $x_1=x_2=\ldots=x_n$.
Following the tradition established in field theory \cite{BSh}, we call such
distributions the quasilocal distributions. $E_n^{(A)_n}$ depend locally on
the field $\Gamma$, so the variational derivatives
$\delta_{A_{n+1}}...\delta_{A_{n+m}}E_n^{(A)_n}$
are quasilocal distributions with respect to all space arguments
$x_1,...,x_n,x_{n+1},...,x_{n+m}$. $E_n^{(A)_n}$
have the following symmetry properties with respect to the simultaneous
permutations of indices and coordinates
$$
E_n^{A_1...A_kA_{k+1}...A_n}(x_1,\ldots,x_k,x_{k+1},\ldots,x_n;\Gamma)=
$$
$$
=(-1)^{\varepsilon_{A_k}\varepsilon_{A_{k+1}}} E_n^{A_1...A_{k+1}A_k...A_n}
(x_1,\ldots,x_{k+1},x_k,\ldots,x_n;\Gamma),
$$
and the Grassmann parity
$$
\varepsilon(E_n^{(A)_n})=\varepsilon(\nabla_n)+\sum_1^n\varepsilon_{A_k}.
$$
$\nabla_n$ defines the operator on the functions  $f(S)$ as follows:
\begin{equation}\label{5}
(\nabla_nf)(S)=E_n^{(A)_n}(\delta_A\circ S_i)^n
f_,{}_{\overleftarrow{(i)_n}}(S)
\equiv S_{\nabla_n(i)_n}f_,{}_{\overleftarrow{(i)_n}}(S),
\end{equation}
where $S_{\nabla_n(i)_n}=S_{\nabla_n(i)_n}[S]={1\over
n!}\nabla_n(S_i)^n=E_n^{(A)_n} (\delta_A\circ S_i)^n$ is
the derivative from
$S_{i_1}$, \ldots, $S_{i_n}$
{\bf local} functional with the united (multi)index
``$\nabla_n(i)_n$'', $\varepsilon(S_{\nabla_n(i)_n})=
\varepsilon(\nabla_n)+\sum_{k=1}^n\varepsilon(i_k)$.
It is evident that  $(\nabla_nf)(S)$ is a functional of the type considered.
Definition (\ref{5}) can obviously be extended to the case where $S_i$ are
not necessarily local.
The general local differential operator $\nabla$ is a
linear combination of the operators $\nabla_n$
$$
\nabla=\sum_{n=0}c_n\nabla_n,
$$
$c_n$  are constants (independent of $x$ and $\Gamma$). $\nabla_0=S_0$ is
an arbitrary local functional: formula (\ref{5}) enables us to consider
the multiplication by a local functional as the zero order differential
operator. Below, we extend this definition to infinite sums and, in such a
case, operate with formal series postponing the question on the sense of
series convergence.

The general form of the operators that are "well" defined on the considered
space of multilocal functionals is a smooth function
$\omega(\nabla)=\omega(\nabla_{n_1},\nabla_{n_2},\ldots)$
of the local operators $\nabla_n$; we call them the multilocal operators.
We assume the functions $\omega$ are given by formal series that also
defines the rule of the action of multilocal operators on multilocal
functionals.  It is evident that the image of a multilocal operator belongs
to the space of multilocal functionals.

Let us now return to the definition of the operators $\nabla_n$. The given
definition of the operator $\nabla_n$ differs from its canonical
understanding, as the differential operator that satisfies Leibnitz rule and
is then denoted by $\Delta_n$ (below, we call such operators the canonical
differential operators), by the additional requirement that it is forbidden
for two or more derivatives $\delta/\delta\Gamma^A(x)$ at the same
point $x$ to act on one and the same local functional. This requirement is
equivalent to the rule
$\delta/\delta\Gamma^{A_1}(x)\delta/\delta\Gamma^{A_2}(x)S(\Gamma)$ $=$ $0$,
i.e. the prescription $\delta(0)=0$, in particular,
$$
\nabla_n(S_i)^m=0,\quad n\ge m+1.
$$
The relationship between the operators $\nabla_n$ and $\Delta_n$ can also be
explained from the quasiclassical expansion point of view. It was noted long
ago that the emergence of $\delta(0)$ in quantum field theory is accompanied
by additional smallness in powers of the Plank constant $\hbar$, the parameter
of the quasiclassical expansion (the loop expansion).  Therefore, the
operator $\nabla_n$ corresponding to differential expression
(ref{3}) can be
considered as a leading quasiclassical approximation to the canonical
differential operator $\Delta_n$. The exact meaning of this statement is the
following. Let us make the substitutions $ S_i\to S_i/\hbar$, $f(S)\to
f(S/\hbar)$ in functionals, $\delta/\delta\Gamma\to
\hbar\delta/\delta\Gamma$ in derivatives with respect to the fields, and, at
last, $\nabla_n\to \nabla_n(\hbar)=(1/\hbar) E_n^{(A)_n}(\hbar\delta_A)^n$ in
differential expression (\ref{3}).

In order that the equations to follow might have a sense,
it is necessary to refuse from locality:
of either the differential operators, by regularizing their coefficient
functions
$E_n^{(A)_n}(x_1,...,x_n)$, or functionals, by assuming $S_i$ to
be regularized nonlocal functionals (naturally, we can assume this and
that). Then it is easy to see that we have for the action of the canonical
differential operator
$$
\Delta_n(\hbar)f(S/\hbar)=\left(\frac{1}{\hbar} (S_{\nabla_n(i)_n})\right)
f_,{}_{\overleftarrow{(i)_n}}(S/\hbar)+ $$ $$
+\sum_{k=1}^n\hbar^k\left(\frac{1}{\hbar}S_{\nabla_n(i)_{n-k}}
\right)
f_,{}_{\overleftarrow{(i)_{n-k}}}(S/\hbar),
$$
where functionals $S_{\nabla_n(i)_{n-k}}$ are given by
$$
S_{\nabla_n(i)_{n-k}}=
E_n^{(A)_k(B)_{n-k}}(\delta_A)^k(\delta_B\circ
S_i)^{n-k},
$$
such that
$$
\Delta_n(\hbar)f(S/\hbar)=\nabla_n(\hbar)f(S/\hbar)+
O(\hbar,S/\hbar),
$$
and  $O(\hbar,S/\hbar)$ as a function of $S/\hbar$ has a nonzero order of
smallness in $\hbar$. When removing the regularization, the functionals
$S_{\nabla_n(i)_{n-k}}$, $k\ge1$ become singularities $\sim \delta(0)$ and so
does $O(\hbar,S/\hbar)$.

It is convenient to set in correspondence to every operator $\nabla_n$ its
symbol $\tilde{\nabla}_n=\tilde{\nabla}_n(\Gamma,p)$, the functional of the
fields $\Gamma=\{\Gamma^A(x)\}$ and their conjugate variables $p=\{p_A(x)\}$,
$\varepsilon(p_A)=\varepsilon(\Gamma^A)=\varepsilon_A$, that is obtained from
differential expression
(\ref{3}) by formally substituting the variable
$p_A(x)$ for the variational derivative symbol $\delta/\delta\Gamma^A(x)$:
$$
\tilde{\nabla}_n=\int dx_1\ldots dx_nE_n^{A_1...A_n}(x_1,\ldots,x_n;
\Gamma)p_{A_1}(x_1)...  p_{A_n}(x_n)=E_n^{(A)_n}(p_A)^n.
$$

\subsection{Operator algebra}

The considered space of functionals is closed under the action of the
multilocal operators $\omega(\nabla)$, so those form an associative algebra.
The algebra of the local operators as an associative algebra is not closed
in itself: the product of two operators $\nabla_n\nabla_m$
is not a local differential operator of the $\nabla$-type. Really, the
result of the action of the (bilocal) operator $\nabla_n\nabla_m$ on a
 functional $f(S)$ contains second order variational derivatives of local
functionals $S_i$ at different points, what is impossible for the action of
a local operator of the $\nabla$-type.

However, as the Lie algebra, the algebra of the $\nabla$-operators proves to
be closed: if $\nabla^{(1)}$, $\nabla^{(2)}$ are local, then
their commutator
$$
[\nabla^{(1)},\nabla^{(2)}]=\nabla^{(1)}\nabla^{(2)}-
(-1)^{\varepsilon(\nabla^{(1)})\varepsilon(\nabla^{(2)})}\nabla^{(2)}
\nabla^{(1)}
$$
is also a local differential operator. In other words, we may say that the
space of the local operators is closed under the $ad$-operation,
\begin{equation}\label{6}
ad\nabla^{(1)}(\nabla^{(2)})=[\nabla^{(1)},\nabla^{(2)}]=\nabla^{(3)}.
\end{equation}
It is sufficient to verify this fact for the homogeneous operators
$\nabla_n$. If
$$
\nabla_m=E_m^{(A)_m}(\delta_A)^m,\quad\nabla_n=E_n^{(A)_n}(\delta_A)^n
$$
are local, then
$$
[\nabla_m,\nabla_n]=
mE_m^{C(A)_{m-1}}\delta_CE_n^{(B)_n}(-1)^{\varepsilon(E_n^{(B)_n})
\sum_{k=1}^{m-1}\varepsilon_{A_k}}(\delta_A)^{m-1}(\delta_B)^n-
$$
$$
-(-1)^{\varepsilon(\nabla_m)\varepsilon(\nabla_n)}nE_n^{C(A)_{n-1}}
\delta_CE_m^{(B)_m}(-1)^{\varepsilon(E_m^{(B)_m})
\sum_{k=1}^{n-1}\varepsilon_{A_k}}(\delta_A)^{n-1}(\delta_B)^m\equiv
$$
$$
\equiv E_{m+n-1}^{([m,n])(A)_{m+n-1}}(\delta_A)^{m+n-1}
=\nabla^{([m,n])}_{m+n-1}.
$$
is  local too.
Really, the coefficient functions $E_{m+n-1}^{([m,n])(A)_{m+n-1}}$
defined by the last but one equality are quasilocal distributions in all
coordinates $x_1,..., x_{n+m-1}$ (because,in fact, they are convolutions of
quasilocal distributions); they are defined uniquely in the class of
distributions that are symmetric with respect to simultaneous permutations
of coordinates and indices. This fact is simply formulated in terms of
the $\tilde{\nabla}$-symbols. Let us consider the Lie algebra
 of functionals $F(\Gamma,p)$ generated by the canonical Poison bracket:  if
 $F=F(\Gamma,p)$, $G=G(\Gamma,p)$, then
$$
\{F,G\}={\delta_rF\over\delta
p_A}{\delta G\over\delta\Gamma^A}-
(-1)^{\varepsilon(F)\varepsilon(G)}{\delta_rG\over\delta p_A}
{\delta F\over\delta\Gamma^A},
$$
here and below, the index $r$  denotes the right
derivative.  $\Gamma$ and $p$ are conjugate just in the sense of this
bracket:
$$
\{p_A(x_1),\Gamma^B(x_2)\}=\delta_A^B\delta(x_1-x_2).
$$
Then it is easy to verify that if $\tilde{\nabla}_m$, $\tilde{\nabla}_n$
are the symbols of $\nabla_m$ and $\nabla_n$ respectively,
$\tilde{\nabla}^{([m,n])}_{m+n-1}$ is the symbol of their commutator, then
$$
\tilde{\nabla}^{([m,n])}_{m+n-1}=\{\tilde{\nabla}_m, \tilde{\nabla}_n\}.
$$
On any $\nabla$, this result extends by linearity.

\section{Operators $\exp\nabla$ and functionals $\exp S$ }

Later on, we need the multilocal operators of the form $\exp\nabla$ treated
as formal series
$$
e^\nabla=\sum_{n=0}\frac{1}{n!}\nabla^n.
$$
The significance of these operators is determined by the fact that many
transformations in field theory, in particular, a change of variables
$\Gamma\to\Gamma^\prime$ in functionals, are formulated in their terms.
Let us turn our attention to the properties of these operators following
from the fact proved just now that the Lie algebra of the $\nabla$-operators
is closed, or equivalently, that the set of these operators is invariant
under the $ad$-operation: if $\nabla^{(1)}$, $\nabla^{(2)}$ are local
operators, then $ad\nabla^{(1)}(\nabla^{(2)})$ is also a local differential
operator. Then the same is valid for
$$
(ad\nabla^{(1)})^n(\nabla^{(2)})=
\underbrace{[\nabla^{(1)},[\nabla^{(1)},\ldots,[\nabla^{(1)},\nabla^{(2)}]
\ldots]]}_n,\quad n=0,1,2,\ldots,
$$
the proof is evident by induction, and, at last, for
\begin{equation}\label{7} Ad(e^{\nabla^{(1)}})(\nabla^{(2)})\equiv
e^{\nabla^{(1)}}\nabla^{(2)}e^{-\nabla^{(1)}}=
e^{ad\nabla^{(1)}}(\nabla^{(2)})=\sum_{n=0}\frac{1}{n!}
(ad\nabla^{(1)})^n(\nabla^{(2)})=\nabla^{(3)},
\end{equation}
$\nabla^{(3)}$ is a local operator. In other words, the space of local
differential operators is invariant under the $Ad$-operation too. In addition,
it is true that
\begin{equation}\label{8}
e^{\nabla^{(1)}}e^{\nabla^{(2)}}=e^{\nabla^{\prime(3)}},
\end{equation}
where $\nabla^{(3)}$ is a local differential operator. This fact is a direct
consequence of the Baker-Hausdorff-Dynkin formula \cite{BHD}:
$$
\nabla^{\prime (3)}=\nabla^{(1)}+\nabla^{(2)}+\frac{1}{2}[\nabla^{(1)},
\nabla^{(2)}]+...,
$$
the dots stand for a series in repeated commutators of $\nabla^{(1)}$  and
$\nabla^{(2)}$.

In other words, the set of the operators  $\exp\nabla$ is invariant under
associative multiplication (and defines the corresponding associative algebra
in a natural way ).

The functionals of the form $f(S)=\exp S$, where  $S$ is a local functional of
the form (\ref{1}), play a special role in quantum theory. Below, we restrict
ourselves to the case of the Bose-functionals, $\varepsilon(S)=0$.
For such functionals, the action of a local operator
$\nabla=\sum_{n=0}c_n\nabla_n$ reduces to the multiplication by a local
functional:
\begin{equation}\label{9} \nabla
e^S=S_{\nabla}[S]e^S,\quad S_{\nabla}[S]=\sum_{n=0}c_n\frac{1}{n!}
\nabla_nS^n=\sum_{n=0}c_nE_n^{(A)n}(\delta_A S)^n.
\end{equation}
The set of the functionals of this type proves to be invariant under the
action of the operators $\exp{\nabla}$
\begin{equation}\label{10}
e^{\nabla}e^S=e^{S^\prime},
\end{equation}
where $S^\prime$ is a local functional. To prove this, let us consider
the one-parameter family of the operators $\exp{(\alpha\nabla)}$, where
$0\le\alpha\le1 $. Let us write
\begin{equation}\label{11}
e^{\alpha\nabla}e^{S}=e^{S^\prime(\alpha)},
\end{equation}
where the functional $S^\prime(\alpha)$  depends smoothly on
$\alpha$,  $S^\prime(0)=S$, $S^\prime(1)=S^\prime$
\begin{equation} \label{12}
S^\prime(\alpha)=S+\alpha S^{(1)}+...= \sum_{m=0}\alpha^mS^{(m)},\quad
S^{(0)}=S.
\end{equation}
For this functional, a differential equation naturally arises: differentiating
Eq(\ref{11}) with respect to $\alpha$, we obtain
$$
\nabla e^{S^\prime(\alpha)}=\frac{\partial S^\prime(\alpha)}{\partial\alpha}
e^{S^\prime(\alpha)}.
$$
But according to Eq(\ref{9}) that holds true also for functionals $S$ not
necessarily local,
$$
\nabla
e^{S^\prime(\alpha)}=S_{\nabla}[S^\prime(\alpha)]e^{S^\prime(\alpha)}, $$ $$
S_{\nabla}[S^\prime(\alpha)]=\sum_{n=0}c_nE_n^{(A)_n} (\delta_A
S^\prime(\alpha))^n,
$$
so we finally obtain the differential equation for $S^\prime_{\nabla}(\alpha)$
\begin{equation}\label{13}
\frac{\partial S^\prime(\alpha)}{\partial\alpha}=S_\nabla[S^\prime(\alpha)]
\end{equation}
with the initial condition
\begin{equation}\label{14}
S^\prime(0)=S.
\end{equation}
In the class of functionals smooth in $\alpha$, the solution of
Eq(\ref{13},{14}) is unique.  Namely, substituting expansion (\ref{12}) into
Eq(\ref{13}), we obtain the recurrent relations for $S^{(m)}$.  In particular,
$$
S^{(1)}=S_{\nabla}[S].
$$
It is interesting to note that the operation  $\exp\nabla$
is in agreement with the quasiclassical expansion in the following sense.
Let us make the substitutions
$$
\nabla\to\nabla(\hbar)=
\sum_{n=0}c_n\nabla_n(\hbar)=
\sum_{n=0}c_n\left({1\over\hbar}E_n^{(A)_n}(\hbar\delta_A)^n\right),\quad
S\to{1\over\hbar}S.
$$
Consider a triple of operators $\nabla^{(i)}, i=1, 2, 3$. Then it turns out
that if $\nabla^{(i)}$ are related by Eq(\ref{6}), the same holds true for
$\nabla^{(i)}(\hbar)$
$$
[\nabla^{(1)},\nabla^{(2)}]=\nabla^{(3)}\quad\to\quad
[\nabla^{(1)}(\hbar),\nabla^{(2)}(\hbar)]=\nabla^{(3)}(\hbar).
$$
It is sufficient to verify this for the homogenous operators $\nabla_n$.
As a consequence, if $\nabla^{(i)}$ are related by Eq(\ref{7}) or Eq(\ref{8}),
the same holds true for $\nabla^{(i)}(\hbar)$
$$
e^{\nabla^{(1)}}\nabla^{(2)}e^{-\nabla^{(1)}}=
\nabla^{(3)}\quad\to\quad
e^{\nabla^{(1)}(\hbar)}\nabla^{(2)}(\hbar)e^{-\nabla^{(1)}(\hbar)}=
\nabla^{(3)}(\hbar),
$$
$$
e^{\nabla^{(1)}}e^{\nabla^{(2)}}=e^{\nabla^{(3)}}\quad\to\quad
e^{\nabla^{(1)}(\hbar)}e^{\nabla^{(2)}(\hbar)}=e^{\nabla^{(3)}(\hbar)}.
$$
It is easy to verify also that if an operator $\nabla$, functionals $S$ and
$S_{\nabla}[S]$ are related by Eq(\ref{9}), the same holds true for
$\nabla(\hbar)$, $S/\hbar$, $S_\nabla/\hbar$ (with the same $S_\nabla$)
$$
\nabla(\hbar)e^{{1\over\hbar}S}={1\over\hbar}S_{\nabla}[S]
e^{{1\over\hbar}S}.
$$
Finally, if an operator $\nabla$, functionals $S$
and $S^\prime$ are related by Eq(\ref{10}), then it is true that
$$
e^{\nabla(\hbar)}e^{{1\over\hbar}S}=e^{{1\over\hbar}S^\prime},
$$
$S^\prime$ being independent of $\hbar$.
The proof repeats the proof of Eq(\ref{10}):
$\hbar$  happily cancels in the equation for $S^\prime(\alpha)$
that coincides with Eq(\ref{13}), (\ref{14}). The solution
$S^\prime(\alpha)$ does not depend on $\hbar$.

\section{Change of variables}

Multilocal functionals $\Phi(\Gamma)$ and multilocal differential  operators
$\omega(\nabla)$ admit local change of variables
$$
\Gamma^A=\Gamma^A(x)\quad \to\quad \Gamma^{\prime A}=
\Gamma^{\prime A}(x,\Gamma(x),\partial_\mu\Gamma(x),\ldots),
$$
where $\Gamma^{\prime A}(x,\Gamma(x),\partial_\mu\Gamma(x),..)$
are smooth functions of fields $\Gamma(x)$
and their finite order derivatives.
This means that $\delta\Gamma^{\prime A}(x)/\delta\Gamma^B(y)$
are quasilocal distributions
$$
\frac{\delta\Gamma^{\prime A}(x)}{\delta\Gamma^B(y)}
=P^A_B(x,\Gamma(x),\partial_\mu\Gamma(x),\ldots,\partial_\mu)\delta(x-y),
$$
$P^A_B(x,\Gamma(x),\partial_\mu\Gamma(x),\ldots,\partial_\mu)$
is a polynomial in  $\partial_\mu$ with coefficients that are local
functions of the fields $\Gamma(x)$ and their derivatives. We assume
that the change of variables conserves the Grassmann parity
$\varepsilon(\Gamma^{\prime A})=\varepsilon(\Gamma^A)$. We also assume
that the change is invertible (at least perturbatively in powers of the field
 derivatives), the inverse change of variables
$$
\Gamma^{\prime A}\quad\to\quad\Gamma =
\Gamma^{A}(x,\Gamma^\prime(x),
\partial_\mu\Gamma^\prime(x),..)
$$
being also local
$$
\frac{\delta\Gamma^{A}(x)}{\delta\Gamma^{\prime B}(y)}
Q^A_B(x,\Gamma(x),\partial_\mu\Gamma(x),\ldots,\partial_\mu)\delta(x-y).
$$
Hence, the local changes of variables form a group (of diffeomorphisms),
at least in some neighborhood of the point $\Gamma^A=0$.

The local change of variables defines the local transformation of functionals
in a natural way.

We assume that the functionals $\Phi(\Gamma)$ transform according to the
scalar (anti)repre\-sentation of the diffeomorphism group
$$
\Phi^\prime(\Gamma)=\Phi(\Gamma^\prime).
$$
It is evident that if  $S(\Gamma)$ is a local functional of the form
(\ref{1}), then $S^\prime(\Gamma)$ is also local. For multilocal functionals
of general form (\ref{2}) we have
$$
(f^\prime(S))(\Gamma)=(f(S^\prime))(\Gamma)=(f(S))(\Gamma^\prime).
$$
Thus, the local transformations take the considered set of
functionals into itself.

Correspondingly, the local changes of variables induce
the transformations of local differential operators $\nabla$
\begin{equation}\label{14a}
\nabla_{(\Gamma)}\quad\rightarrow\quad\nabla^\prime_{(\Gamma)}=
\nabla_{(\Gamma^\prime)},
\end{equation}
This transformations are naturally defined as follows:
$$
(\nabla f(S))(\Gamma)=\nabla_{(\Gamma)} f(S(\Gamma))\quad\rightarrow\quad
(\nabla^\prime f^\prime(S))(\Gamma)=
(\nabla f(S))(\Gamma^\prime)=\nabla_{(\Gamma^\prime)}f(S(\Gamma^\prime)).
$$
For differential operator $\nabla_n$ (\ref{3}), (\ref{4}) this transformation
law takes the form
\begin{equation} \label{15}
\nabla_n^\prime=E_n^{\prime(A)_n}(\delta_A)^n,
\end{equation}
where
$$
E_n^{\prime A_1...A_n}(x_1,\ldots,x_n;\Gamma)=\int dy_1\ldots dy_n
(-1)^{\sum_{k=1}^{n-1}\varepsilon_{A_k}\sum_{i=k+1}^n
(\varepsilon_{A_i}+\varepsilon_{B_i})}\times
$$
$$
E_n^{B_1...B_n}(y_1,\ldots,y_n;\Gamma^\prime)
\frac{\delta\Gamma^{A_1}(x_1)}{\delta\Gamma^{\prime B_1}(y_1)}...
\frac{\delta\Gamma^{A_n}(x_n)}{\delta\Gamma^{\prime B_n}(y_n)}.
$$
The coefficient functions $E_n^{A_1...A_n}(x_1,\ldots,x_n;\Gamma)$
transform according to the tensor (anti)representation. It is evident
that if $E_n^{A_1...A_n}(x_1,\ldots,x_n;\Gamma)$  is a quasilocal
distribution, then $E_n^{^\prime A_1...A_n}(x_1,\ldots,x_n;\Gamma)$ is a
quasilocal distribution too (as a convolution of quasilocal distributions).
Therefore, the operator $\nabla^\prime_n$ is local. For arbitrary operators
$\nabla=\sum c_n\nabla_n$, the rule for change of variables extends by
linearity.  Finally, for multilocal operators $\omega(\nabla)$, the rule for
change of variables has the form
$$
\omega(\nabla)\quad\rightarrow\quad\omega(\nabla^\prime).
$$

It is evident that the analogous properties hold true if we define the
transformations of the functionals and the differential operators under
the change of variables according to the scalar representation
$$
\Phi^\prime(\Gamma^\prime)=\Phi(\Gamma),\quad
\nabla^\prime_{(\Gamma^\prime)}=\nabla_{(\Gamma)}.
$$

The local changes of variables are realized by operators $\exp{\nabla_1}$.
Namely, the following formulas take place:
$$
\Gamma^{\prime A}(x)=e^{\nabla_1}\Gamma^{A}(x)
$$
with some local differential operator $\nabla_1$,
$$
S^\prime(\Gamma)=S(\Gamma^\prime)=e^{\nabla_1}S(\Gamma),
$$
$$
e^{\nabla_1}f(S)=f(S^\prime).
$$
If $\Phi(\Gamma)$ is considered as a multiplication operator in the space of
functionals, then
$$
e^{\nabla_1}\Phi(\Gamma)e^{-\nabla_1}=\Phi^\prime(\Gamma)=
\Phi(\Gamma^\prime),
$$
whereas
$$
e^{\nabla_1}\frac{\delta}{\delta\Gamma^A(x)}e^{-\nabla_1}=
\frac{\delta}{\delta\Gamma^{\prime A}(x)}=
\int dy\frac{\delta\Gamma^B(y)}
{\delta\Gamma^{\prime A}(x)}\frac{\delta}{\delta\Gamma^B(y)},
$$
correspondingly
$$
e^{\nabla_1}\nabla e^{-\nabla_1}=\nabla^\prime.
$$

\section{Antibracket, master equation, G transformations}

\subsection{Antibracket, canonical transformations}

Let $\nabla_2$ be a local differential operator of the second order
\begin{equation}\label{16}
\nabla_2=\frac{1}{2}\int
dx_1dx_2E_2^{A_1A_2}(x_1,x_2;\Gamma) \frac{\delta}{\delta\Gamma^{A_1}(x_1)}
\frac{\delta}{\delta\Gamma^{A_2}(x_2)},
\end{equation}
with Grassmann parity equal to unity, $\varepsilon(\nabla_2)=1$, and
nilpotent,
\begin{equation}\label{17}
\nabla_2^2=\frac{1}{2}[\nabla_2,\nabla_2]=0.
\end{equation}
This operator defines a bilinear operation on local functionals
that is called antibracket. Namely, let $S_1$, $S_2$ be arbitrary local
functionals, then the  antibracket $(S_1,S_2)$ is
defined as
\begin{equation}\label{18}
\begin{array}{c}
(S_1,S_2)=(-1)^{\varepsilon(S_1)}\nabla_2(S_1S_2)=
2(-1)^{\varepsilon(S_1)}S_{\nabla_2 12}=
\\
\displaystyle=\int dx_1dx_2E_2^{A_1A_2}(x_1, x_2;\Gamma)
(-1)^{(\varepsilon_{A_2}+1)\varepsilon(S_1)}
\frac{\delta S_1}{\delta\Gamma^{A_1}(x_1)}
\frac{\delta S_2}{\delta\Gamma^{A_2}(x_2)},\\
\displaystyle
\varepsilon((S_1,S_2))=\varepsilon(S_1)+\varepsilon(S_2)+1,
\end{array}
\end{equation}
the antibracket is evidently a local functional.
We can now write down (see (\ref{5}))
\begin{equation}\label{18a}
\nabla_2 f(S)={1\over2}(-1)^{\varepsilon(S_{i_1})}(S_{i_1},
S_{i_2})f_{,i_{2}i_{1}}(S).
\end{equation}
It is convenient to represent the antibracket in the following form:
$$
(S_1, S_2)=\int dx_1 dx_2 \frac{\delta_r
S_1}{\delta\Gamma^{A_1}(x_1)} {\tilde{E}}_2^{A_1A_2}(x_1,x_2;\Gamma)
\frac{\delta S_2}{\delta\Gamma^{A_2}(x_2)},
$$
where the metric tensor  ${\tilde{E}}_2^{A_1A_2}$ of the antibracket is
\begin{equation}\label{19}
{\tilde{E}}_2^{A_1A_2}=(-1)^{\varepsilon_{A_1}(\varepsilon_{A_2}+1)}
E_2^{A_1A_2}=-(-1)^{(\varepsilon_{A_1}+1)(\varepsilon_{A_2}+1)}
{\tilde{E}}_2^{A_2A_1}.
\end{equation}
The antibracket evidently has the antisymmetry property
$$
(S_1, S_2)=-(-1)^{(\varepsilon(S_1)+1)(\varepsilon(S_2)+1)}(S_2, S_1).
$$
As it follows from  (\ref{18}), (\ref{18a}), the nilpotency property
(\ref{17}) is equivalent to the Jacoby identity for the antibracket
\begin{equation}\label{20}
\nabla^2_2(S_1S_2S_3)\equiv0\quad\to\quad
(-1)^{(\varepsilon(S_1)+1)(\varepsilon(S_3)+1)}(S_1,(S_2,S_3))+
\hbox{cycle}(1,2,3)=0.
\end{equation}
In terms of the coefficient functions ${\tilde{E}}_2^{A_1A_2}$
this condition means (it is sufficient to take
$S_1=\Gamma^A(x)$, $S_2=\Gamma^B(y)$, $S_3=\Gamma^C(z)$ in (\ref{20}))
$$
(-1)^{(\varepsilon_A+1)(\varepsilon_C+1)}
\int du\tilde{E}_2^{AD}(x,u;\Gamma)\frac{\delta}{\delta\Gamma^D(u)}
\tilde{E}_2^{BC}(y,z;\Gamma)+\hbox{cycle}(Ax,By,Cz)=0,
$$
or, in the condensed notations,
\begin{equation}\label{21}
(-1)^{(\varepsilon_A+1)(\varepsilon_C+1)}\tilde{E}_2^{AD}\delta_D
\tilde{E}_2^{BC}+\hbox{cycle}(A,B,C)=0.
\end{equation}
Of course, the converse is also true: if the tensor $\tilde{E}_2^{A_1A_2}$
satisfies antisymmetry condition (\ref{19}) and Jacoby identity (\ref{21}),
then the operator $\nabla_2$ (\ref{16}) with  $E_2^{A_1A_2}$ defined by
(\ref{19}) is nilpotent.
It is a standard to assume that under the change of variables $\Gamma$ $\to$
$\Gamma^\prime$, the antibracket  transforms, $(., .)$ $\to$ $(., .)^\prime$
(the index``$^\prime$'' at the bracket is the symbol of a new
antibracket), according to the following rule:
$$
(S_1,S_2)(\Gamma)
\quad\rightarrow\quad(S_1^\prime,S_2^\prime)^\prime(\Gamma)=
\delta_{rA}S_1^\prime\tilde{E}_2^{\prime AB}\delta_BS_2^\prime(\Gamma)=
(S_1,S_2)(\Gamma^\prime),
$$
wherefrom it follows that the metric tensor of the antibracket transforms
according to the tensor (anti)representation:
$$
\tilde{E}_2^{\prime A_1A_2}(x_1,\ldots, x_n;\Gamma)=\int dy_1dy_2
\frac{\delta_r\Gamma^{A_1}(x_1)}{\delta\Gamma^{\prime B_1}(y_1)}
\tilde{E}_2^{B_1B_2}(y_1, y_2;\Gamma^\prime)
\frac{\delta\Gamma^{A_2}(x_2)}{\delta\Gamma^{\prime B_2}(y_2)}.
$$
This transformation properties of the antibracket are in agreement with the
rules that are obtained from definition (\ref{18}) of the antibracket:
$(S_1,S_2)=(-1)^{\varepsilon(S_1)}\nabla_2(S_1S_2)$,
$(S_1,S_2)^\prime=(-1)^{\varepsilon(S_1)}\nabla^\prime_2(S_1S_2)$,
and transformation properties (\ref{14a}) of the operators $\nabla$.

In connection with the antibracket (as well as with the master equation,
see below), the question arises on the field transformation
$\Gamma$ $\rightarrow$ $\Gamma^\prime=e^{\nabla_1}\Gamma$
that conserve the antibracket (local canonical changes of variables)
$$
(S_1, S_2)^\prime(\Gamma)=(S_1, S_2)(\Gamma).
$$
In terms of operators, this means
\begin{equation}\label{22}
e^{\nabla_1}\nabla_2e^{-\nabla_1}=\nabla_2,
\end{equation}
or
$$
[e^{\nabla_1},\nabla_2]=0,
$$
i.e. the transformation conserves the operator $\nabla_2$.
In terms of the coefficient functions, this means
$$
\tilde E_2^{\prime A_1A_2}(x_1,x_2;\Gamma)=
\tilde E_2^{A_1A_2}(x_1,x_2;\Gamma).
$$
The class of such transformations depends essentially on the properties of
the coefficient functions $\tilde E_2^{A_1A_2}$ (for example, in the trivial
case ${\tilde{E}}_2^{A_1A_2}=0$, it includes all transformations).  In the
finite-dimension case, where the coefficient functions are finite-dimensional
matrices $E^{ij}_2$, this class of transformations are determined by
the rank of $E^{ij}_2$.

It is easy to see that property (\ref{22}) is satisfied if
$$
\nabla_1=[\nabla_2,F]=\nabla_2F+F\nabla_2,
=-\int dx_1dx_2\frac{\delta_rF}{\delta\Gamma^{A_1}(x_1)}
{\tilde{E}}_2^{A_1A_2}(x_1, x_2;\Gamma)
\frac{\delta}{\delta\Gamma^{A_2}(x_2)},
$$
where $F$ is a local fermion functional, $\varepsilon(F)=1$.
This follows from the relation
$$
[\nabla_1,\nabla_2]=0
$$
that is easy to verify. It is obvious that
$$
\nabla_1S=(S,F).
$$
Thus, the transformation
\begin{equation}\label{23}
\Gamma^A\to\Gamma^{\prime A}=e^{[\nabla_2, F]}\Gamma^A
\end{equation}
is canonical.

In the finite-dimensional case, we can show that if the matrix
$E^{ij}_2$ is nondegenerate, then the transformations generated by
$\exp{[\nabla_2, F]}$ cover all transformations conserving
the antibracket or, what is the same, $\nabla_2$.

\subsection{Master equation, G transformations}

The nilpotent local operator $\nabla_2$,
$\varepsilon(\nabla_2)=1$ defines the master equation for local functionals
$S(\Gamma)$
$$
\nabla_2e^{{i\over\hbar}S}=0
$$
or
\begin{equation}\label{24}
(S, S)=\nabla_2 S^2=0.
\end{equation}
This equation is nontrivial only for Bose functionals.
It is evident that this equation is invariant under the canonical
transformations of functionals induced by local canonical
transformations of fields $\Gamma$ that conserve the antibracket.  In
particular, if $S$ is a solution of the master equation, then
\begin{equation}\label{25} S^\prime=e^{[\nabla_2, F]}S,\quad
\varepsilon(F)=1,
\end{equation}
is a solution too.  Transformation
(\ref{25}) can be represented as \begin{equation}\label{26}
e^{{i\over\hbar}S}\quad\to\quad e^{{i\over\hbar}S^\prime} =e^{[\nabla_2, F]}
e^{{i\over\hbar}S}.
\end{equation}
The invariance of the master equation, as well as of the antibracket, under
canonical transformations (\ref{26}) results from the fact that for a
nilpotent operator $\nabla_2$, the operator $[\nabla_2, F] $
commutes with $\nabla_2$. On the other hand, any operator of the form
$[\nabla_2,\nabla]$, with an arbitrary $\nabla$, commutes with $\nabla_2$.
It follows that any transformation of the form
\begin{equation}\label{27}
e^{{i\over\hbar}S}\quad\to\quad
e^{{i\over\hbar}S^\prime}=e^{[\nabla_2(\hbar),\nabla(\hbar)]}
e^{{i\over\hbar}S},
\end{equation}
with an arbitrary local $\nabla$, $\varepsilon(\nabla)=1$,
takes a local solution $S$ of the master equation to a
local solution  $S^\prime$ (independent of $\hbar$).
We call the transformations of the form (\ref{27}) the
gauge  (G) transformations. Note that to
represent G transformations as an operator acting directly on
$S$ is a rather complicated problem. It is worth noting also that the  G
transformation operators $ \exp{[\nabla_2,\nabla]} $ form an associative
algebra.

However, it turns out that for the solutions of the master equation, any G
transformation (\ref{27}) reduces to the canonical transformation (\ref{26}):
$$
e^{{i\over\hbar}S}\quad\to\quad e^{{i\over\hbar}S^\prime} =e^{[\nabla_2,
F]} e^{{i\over\hbar}S},
$$
with some local Fermi functional $F(\Gamma)$. The fact that the functional
$F$ does exist can be proved, for example, as follows. Instead of the
operator $\nabla$,  we consider the operator $\alpha\nabla$.  Then the
relation $$ e^{\alpha[\nabla_2(\hbar),\nabla(\hbar)]}e^{{i\over\hbar}S}=
e^{[\nabla_2,F(\alpha)]}e^{{i\over\hbar}S}
$$
will hold true if, for example, $F(\alpha)$ is a solution of the equation
\begin{equation}\label{28} \int_0^1d\beta
e^{\beta[\nabla_2,F(\alpha)]}{\partial
F(\alpha)\over\partial\alpha}=S_{\nabla}(\alpha),\quad F(0)=0,
\end{equation}
and the local independent of $\hbar$ Fermi functional
$S_{\nabla}(\alpha)$ is defined by
$$
\hbar\nabla(\hbar)e^{\alpha[\nabla_2(\hbar),\nabla(\hbar)]}
e^{{i\over\hbar}S}= S_{\nabla}(\alpha)e^{\alpha[\nabla_2(\hbar),
\nabla(\hbar)]}e^{{i\over\hbar}S}.
$$
The formal perturbative in $\alpha$ local solution of Eq.(\ref{28}) for
$F(\alpha)$ exists.

The answer to the question whether G transformations act on the solutions
transitively is determined by the properties of the operator $\nabla_2$, i.e.
of its coefficient functions $E_2^{A_1A_2}$, and by additional conditions on
$S$ (it is clear that $S$=0 is a solution anyway).

In the framework of the standard BV  formulation of gauge theories, the
metric $\tilde{E}_2^{A_1A_2}$ is nonsingular whereas the fields $\Gamma$ are
Darboux coordinates for the metric
$$
\Gamma^A(x)=\{\Phi^a(x),\Phi^*_a(x)\},\quad\varepsilon(\Phi^*_a)=
\varepsilon(\Phi^a)+1,
$$
\begin{equation}\label{29}
\nabla_2=\int dx(-1)^{\varepsilon(\Phi^a)}
{\delta\over\delta\Phi^a(x)}{\delta\over\delta\Phi^*_a(x)}.
\end{equation}
The fields  $\Phi^a$ are divided into two groups: $\Phi^a$ $=$
$\{\varphi^i,C^\alpha\}$, where  $\varphi^i$  are the fields of an
initial classical theory, $C^\alpha$ are ghost fields.
The fields $\Phi^*_a$ (called antifields) are divided  analogously.
The ghost number $gh$ is ascribed to all variables: $gh(\varphi^i)=0$,
$gh(C^\alpha)=1$, $gh(\Phi^*_a)=-gh(\Phi^a)-1$. The solution $S$ of master
equation (\ref{24}) is sought in the form of the power series expansion in $C$:
\begin{equation}\label{30}
S={\cal S}+ \sum_{n=1}S^{(n)},\quad
S^{(n)}=O(C^n), \quad\varepsilon(S)=gh(S)=0,
\end{equation}
${\cal S}$ is a local action of the classical theory. The action $\cal S$
is assumed to have  a gauge symmetry, i.e. satisfies the gauge identities:
$$
R^i_\alpha(x,\varphi(x),
\partial_\mu\varphi(x),\ldots,\partial_\mu){\delta{\cal
S}\over\delta\varphi^i(x)}=0,
$$
where $R^i_\alpha(x,\varphi(x), \partial_\mu\varphi(x),\ldots,\partial_\mu)$
is a polynomial in  $\partial_\mu$ with local in $x$ coefficients.
In addition, $\cal S$ satisfies some regularity conditions
(see \cite{Reg}
and references therein), whose essence is that
in an arbitrary set of the extremals $\delta{\cal S}/\delta\varphi^i(x)$
and their space derivatives, it is possible to distinguish linearly
(with local coefficients) dependent and linearly independent elements,
and the only relations between them are linear combinations (with local
coefficients) of the gauge identities and their space derivatives.
Under these regularity conditions, it was proved \cite{Reg} that the master
equation does have  the local solutions of form (\ref{30}) and the
arbitrariness in these solutions is described by  canonical transformations
(\ref{25}), i.e.the canonical transformations act on the space of the
solutions of form (\ref{30}) transitively.

Let us compare the expressions given above  with the analogous
formal expressions in the BV formalism \cite{BV}. The analog of the nilpotent
operator $\nabla_2$ is the canonical second order
differential operator $\Delta_2$  (we restrict ourselves to the case where the
variables $\Gamma^A(x)$ are the Darboux coordinates for the metric
$E_2^{A_1A_2}$, i.e., the differential expressions for $\nabla_2$ and
$\Delta_2$ have form (\ref{29})). Formally, the action in quantum field
theory must satisfy the quantum master equation
$$
\Delta_2e^{{i\over\hbar}S}=0
$$
or
\begin{equation}\label{31}
{1\over2}(S,S)=i\hbar\Delta_2S.
\end{equation}
Equation (\ref{24}) and
formal equation (\ref{31}) differs by the terms proportional to
$\sim\hbar\delta(0)$, i.e. master equation (\ref{24}) is a quasiclassical
approximation to master equation (\ref{31}), and the solutions of master
equation (\ref{24}) are the zero order in $\hbar$ (quasiclassical)
approximations to the solutions of quantum master equation (\ref{31}).

A formal operator of the G transformations is
\begin{equation}\label{32}
e^{[\Delta_2,F]}=De^{[\nabla_2,F]}=e^{{i\over\hbar}(-i\hbar\ln D)}
e^{[\nabla_2,F]},
\end{equation}
where
$$
D={D(\Phi^\prime)\over D(\Phi)}=
\left({D(\Gamma^\prime)\over D(\Gamma)}\right)^{1/2},\quad\Gamma^{\prime A}=
e^{[\nabla_2,F]}\Gamma^A.
$$
A functional $S^\prime$ related to a functional $S$ by
$$
e^{{i\over\hbar}S^\prime}=e^{[\Delta_2,F]}e^{{i\over\hbar}S},
$$
is a solution of the quantum master equation if  $S$ is.
Thus, the operator $\exp{([\nabla_2,F])}$ is a quasiclassical
approximation to formal operator (\ref{32}).

Another fact is also worth noting. As a standard, the following boundary
conditions are imposed on a solution of master equation (\ref{31}):
$$
\left.S\right|_{\hbar=0,C=0}={\cal S}.
$$
In this case, although  operators (\ref{32})
take a solution of the master equation to a solution but do not act
on the solutions transitively \cite{Trans}.

\section{$Sp(2)$ master equation}

In this section, we consider the $Sp(2)$ master equation \cite{BLT}
that describes the $Sp(2)$-symmetric generalization of the BV formalism. The
full set of variables $\Gamma^\Sigma$ of the theory (in this section we
change a little the notations for the indices at the variables $\Gamma$) is
divided into the groups $\Phi^A, \Phi_{Aa}^*, {\bar\Phi}_A,$ $ a, b, c=1,2$,
every variable being a function of the coordinates $x$.
In its turn, the variables $\Phi^A $ are divided into groups
$\Phi^A =(\varphi^i,C^{\alpha b},B^\alpha)$,
where $\varphi^i$ are the classical theory variables, $ C^{\alpha b}$ are the
ghost and antighost fields, and $B^\alpha $ are the gauge introducing fields.
The Grassmann parity is ascribed to all fields:
$ \varepsilon(\Phi^A)=\varepsilon(\bar{\Phi}_A)=\varepsilon_A$,
$ \varepsilon(\Phi^*_{Aa})=\varepsilon_{A}+1$,
$\varepsilon(B^\alpha)=\varepsilon_\alpha$,
$\varepsilon(C^{\alpha b})=\varepsilon_\alpha+1,
$
as well as the new ghost number ngh:
$\hbox{ngh}(\varphi)=0$, $\hbox{ngh}(C^b)=1$, $\hbox{ngh}(B)=2$,
$\hbox{ngh}(\bar{\Phi})=-2-\hbox{ngh}(\Phi)$,
$\hbox{ngh}(\Phi^*_a)=-1-\hbox{ngh}(\Phi)$,
$ \hbox{ngh(FM)}=\hbox{ngh(F)}+\hbox{ngh(M)}$.
In the $Sp(2)$ formalism the effective action
 $ S(\Phi, \Phi_a^*, {\bar\Phi}) $
satisfies  the $Sp(2)$ master equation
\begin{equation}\label{33}
\frac{1}{2}(S,S)^a+\int dx\varepsilon^{ab}\Phi_{Ab}^*(x)
    \frac{\delta}{\delta{\bar\Phi(x)}_{A}}S=0,
\end{equation}
$$
\varepsilon^{ab}=-\varepsilon^{ba},\quad
\varepsilon_{ab}\varepsilon^{bc}=\delta^c_a,
\varepsilon_{12}=\varepsilon^{21}=1,
$$
and the boundary condition
$$
\left.S\right|_{\Phi_{a}^*={\bar\Phi}=0}={\cal S}(\varphi),
$$
where ${\cal S}(\varphi)$ is the initial classical action with
a gauge symmetry,  $ R^i_\alpha \delta {\cal S}/\delta \varphi_i=0$.
In (\ref{33})
$(., .)^a$ denotes the doublet of the antibrackets
$$
(S_1,S_2)^a\equiv\int dx\left( {\delta_rS_1\over\delta\Phi^A(x)}
{\delta S_2\over\delta\Phi^*_{Aa}(x)}-
{\delta_rS_1\over\delta\Phi^*_{Aa}(x)}
{\delta S_2\over\delta\Phi^A(x)}\right).
$$

We introduce the doublet of the operators
$$
\bar{\nabla}^a=\nabla^a_2+\nabla^a_1,\quad\varepsilon(\bar\nabla^a)=1,
$$
$$
\nabla^a_2 = \int dx(-1)^{\varepsilon_{A}}
\frac{\delta}{\delta \Phi^A(x)} \frac {\delta}{\delta\Phi_{Aa}^*(x)},
\quad\varepsilon(\nabla^a_2)=1,
$$
$$
\nabla^a_1=i\int dx\varepsilon^{ab}\Phi_{Ab}^*(x)
    \frac{\delta}{\delta{\bar\Phi}_{A}(x)},\quad\varepsilon(\nabla^a_1)=1.
$$
These operators are nilpotent
$$
\nabla^a_2\nabla^b_2+\nabla^b_2\nabla^a_2=0,\quad
\nabla^a_1\nabla^b_1+\nabla^b_1\nabla^a_1=0,\quad
\nabla^a_2\nabla^b_1+\nabla^b_2\nabla^a_1+
\nabla^a_1\nabla^b_2+\nabla^b_1\nabla^a_2=0,\quad
$$
$$
\bar{\nabla}^a\bar{\nabla}^b+\bar{\nabla}^b\bar{\nabla}^a=0.
$$
The doublet of the antibrackets  is related to the doublet of the
operators $\nabla^a_2$ by $$
(S_1,S_2)^a=(-1)^{\varepsilon(S_1)}\nabla^a_2(S_1S_2).
$$
Using the operators $\bar{\nabla}^a$, we can
 write the $Sp(2)$ master equation in the following equivalent form:
\begin{equation}\label{34}
\bar{\nabla}^a(\hbar)e^{{i\over\hbar}S}=0.
\end{equation}
Under the same assumptions that were described in subsec. 5.2, in \cite {ShaT}
it was shown that the $Sp(2)$ master equation has local solutions,
as well as the arbitrariness in the general solution was found. We shall show
that the formalism considered here is a natural apparatus for this purpose.

We introduce the class of operators, which we call the gauge
(G) transformation operators,
\begin{equation}\label{35} K=e^U,\quad
U={1\over2}\varepsilon_{ab}[\bar{\nabla}^b,[\bar{\nabla}^a,\nabla]],
\quad\varepsilon(U)=0,
\end{equation}
where $\nabla$ is an arbitrary local operator, $\varepsilon(\nabla)=0$.
It is evident that the operator $U$ has the property
\begin{equation}\label{36}
[\bar{\nabla}^a,U]=0,
\end{equation}
and, as a consequence, the G transformation operator commutes with
the operators $\bar{\nabla}^a$
\begin{equation}\label{37}
[\bar{\nabla}^a,K]=[\bar{\nabla}^a,e^U]=0.
\end{equation}
It follows from (\ref{36}) that the operators $U$ form the Lie algebra
$$
[U^{(1)},U^{(2)}]={1\over2}\varepsilon_{ab}[\bar{\nabla}^b,[\bar{\nabla}^a,
\nabla_3]]=U^{(3)},\quad \nabla_3=[U^{(1)},\nabla^{(2)}].
$$
Then, according to the Baker-- Hausdorff--Dynkin formula, we have
$$
e^{U^{(1)}}e^{U^{(2)}}=e^{U^{(3)}},
$$
therefore, the G transformation operators form the associative algebra.
It follows from the results of sec. 3 that if $S$ is a
local functional and independent of
$\hbar$, then the functional $S^\prime$  defined by
$$
e^{{i\over\hbar}S^\prime}=K(\hbar)e^{{i\over\hbar}S},
$$
$$
K(\hbar)=e^{U(\hbar)},\quad U(\hbar)=
{1\over2}\varepsilon_{ab}[\bar{\nabla}^b(\hbar),
[\bar{\nabla}^a(\hbar),\nabla(\hbar)]],
$$
is local and independent of $\hbar$.
In addition, if a functional
 $S$ satisfies $Sp(2)$ master equation (\ref{33}), then because of
(\ref{37}), the functional $S^\prime$ satisfies the $Sp(2)$
master equation too
$$
\bar{\nabla}^a(\hbar)e^{{i\over\hbar}S^\prime}=0.
$$
For the functionals $S$ satisfying the $Sp(2)$ master equation the
relation
$$
e^{{1\over2}\varepsilon_{ab}[\bar{\nabla}^b(\hbar),
[\bar{\nabla}^a(\hbar),\nabla(\hbar)]]}e^{{i\over\hbar}S}=
e^{{1\over2}\varepsilon_{ab}[\bar{\nabla}^b(\hbar),
[\bar{\nabla}^a(\hbar),\nabla_0(\hbar)]]}e^{{i\over\hbar}S}
$$
$$
\nabla_0(\hbar)={1\over\hbar}Y,
$$
holds true, where $Y$ is some local functional that is independent of $\hbar$,
$\varepsilon(Y)=0$.
Indeed, if instead of the operator $\nabla$, we consider the operator
$\alpha \nabla^a$, then we can show that the relation
$$
e^{{i\over\hbar}S^\prime(\alpha)}=
e^{{\alpha\over2}\varepsilon_{ab}[\bar{\nabla}^b(\hbar),
[\bar{\nabla}^a(\hbar),\nabla(\hbar)]]}e^{{i\over\hbar}S}=
e^{{1\over2}\varepsilon_{ab}[\bar{\nabla}^b(\hbar),
[\bar{\nabla}^a(\hbar),{1\over\hbar}Y(\alpha)]]}e^{{i\over\hbar}S}
$$
holds true, for instance, for $Y(\alpha)$ satisfying the equation
$$
\int_0^1d\beta e^{\beta{1\over2}\varepsilon_{ab}[\bar{\nabla}^b(\hbar),
[\bar{\nabla}^a(\hbar),{1\over\hbar}Y(\alpha)]]}
{\partial Y(\alpha)\over\partial\alpha}
e^{-\beta{1\over2}\varepsilon_{ab}[\bar{\nabla}^b(\hbar),
[\bar{\nabla}^a(\hbar),{1\over\hbar}Y(\alpha)]]}e^{{i\over\hbar}
S^\prime(\alpha)}=
$$
$$
=\nabla(\hbar)e^{{i\over\hbar}S^\prime(\alpha)}={i\over\hbar}S_{\nabla}(\alpha)
e^{{i\over\hbar}S^\prime(\alpha)},\quad Y(0)=0,
$$
The perturbative in $\alpha$ solution of the latter equation exists .
Thus, if we consider the action of the G transformation operators (\ref{35})
on the functionals $\exp{({i\over\hbar}S)}$, where $S$ obeys $Sp(2)$ master
equation, we can restrict ourselves to the G transformation operators with
$\nabla=\nabla_0$.

Note that any change of the fields $\Phi^A$,
$$
\Phi^A\quad\rightarrow\Phi^{\prime A}=\Phi^{\prime A}(\Phi)=
e^{T^B(\Phi)\delta_B}\Phi^A,
$$
can be extended to the G transformation in the following sense:
$$
\left.e^{{i\over2}\varepsilon_{ab}[\bar{\nabla}_b,[\bar{\nabla}^a,
\bar{\Phi}_AT^A]]}G(\Gamma)\right|_{\Phi^*_a=\bar{\Phi}=0}=
e^{T^A\delta_A}g(\Phi)=g(\Phi^\prime),
$$
$$
\left.g(\Phi)=G(\Gamma)\right|_{\Phi^*=\bar{\Phi}=0}.
$$

Let us now proceed to describing the arbitrariness in the solution of the
$Sp(2)$ master equation. Let
$S={\cal S}+\sum_{n=1}S_{(n)}$, $S_{(n)}\sim C^kB^{n-k}$ be a solution
of the $Sp(2)$ master equation with
$$ S_{(1)}=\int dx\left(C^{\alpha
a}R^i_\alpha \varphi^*_{ia}+\varepsilon^{ab} C^*_{\alpha
b|a}B^\alpha+B^\alpha
R^i_\alpha\bar{\varphi_i}(-1)^{\varepsilon_i+\varepsilon_\alpha}\right).
$$
It was shown in \cite{ShaT} that if two solutions $S$ and $S_1$
coincide up to the $n$-th order of the series expansion in variables
$C^{\alpha b},B^\alpha$, then their difference $\delta S_{(n+1)}$ in
the $(n+1)$-th order
$$
S_1-S=\delta
S_{(n+1)}+O(C^kB^{n+2-k}),\quad \delta S_{(n+1)}\sim C^kB^{n+1-k},
$$
can be represented in the form
$$
\delta S_{(n+1)}={1\over2}\varepsilon_{ab}\omega^b\omega^aX_{(n+1)},
\quad X_{(n+1)}\sim C^kB^{n+1-k},
$$
$$
\omega^a=\int
dx\left((-1)^{\varepsilon_i}L_i\frac{\delta}{\delta \varphi^*_{ia}} -
(-1)^{\varepsilon_\alpha}R^i_\alpha \varphi^*_{ib}\frac{\delta}{\delta
C^*_{\alpha b|a}} + \right.
$$
$$
+((-1)^{\varepsilon_i}R^i_\alpha\bar{\varphi_i}+\varepsilon^{cb}C^*_{\alpha
b|c}) \frac{\delta}{\delta B^*_{\alpha a}}
\left.
-(-1)^{\varepsilon_\alpha}
\varepsilon^{ab}B^\alpha\frac{\delta}{\delta C^{\alpha b}}
+
\varepsilon^{ab}\Phi^*_{Ab}\frac{\delta}{\delta \bar{\Phi}_A}
\right),
$$
$$
L_i(x)\equiv
\delta {\cal S}/\delta
\varphi^i(x),
$$
$X_{(n+1)}$ is a local functional, $\varepsilon(X_{(n+1)})=0$.
Let us introduce the operators $W^a$
$$
W^a=(S,.\phantom{aa})^a+\int dx
\varepsilon^{ab}\Phi^*_{Ab}\frac{\delta}{\delta \bar{\Phi}_A}.
$$
The following relation is valid
$$
{1\over2}\varepsilon_{ab}\omega^b\omega^aX_{(n+1)}=\left.
{1\over2}\varepsilon_{ab}W^bW^aX_{(n+1)}\right|_{n+1}.
$$
Taking into account (\ref{33}), (\ref{34}), we can directly verify
the validity of the equality
 $$
{i\over\hbar}\left({1\over2}\varepsilon_{ab}W^bW^aX_{(n+1)}\right)
e^{{i\over\hbar}S}=
-{i\over2}\varepsilon_{ab}[\bar{\nabla}^b(\hbar),
[\bar{\nabla}^a(\hbar),{1\over\hbar}X_{(n+1)}]]e^{{i\over\hbar}S}.
$$
With (\ref{37}) taken into account, we obtain that the action $S^\prime_1$,
$$
e^{{i\over\hbar}S^\prime_1}=
e^{{i\over2}\varepsilon_{ab}[\bar{\nabla}^b(\hbar),
[\bar{\nabla}^a(\hbar),{1\over\hbar}X_{(n+1)}]]}e^{{i\over\hbar}S_1},
$$
satisfies the $Sp(2)$ master equation and differs from $S$ starting from
the $(n+2)$-th order. As far as all solutions coincide in the zero order
in  $C$ and $B$, then using the induction method, we finally obtain  the
following result: the general solution $\tilde{S}$ of the  $Sp(2)$ master
equation can be constructed from a particular solution $S$ with the help
of the G transformation operator
$$
e^{{i\over\hbar}\tilde{S}}=K(\hbar)e^{{i\over\hbar}S},
$$
i.e., the G transformations act on the solutions of the $Sp(2)$ master
equation transitively, and we can restrict ourselves to the G transformation
operators of form (\ref{35}) with $\nabla=\nabla_0$.

As in subsec. 5.2, we can verify that the operators $\bar{\nabla}^a$
and the G transformation operators (\ref{35}) are the quasiclassical
approximations to the formal nilpotent operators $\bar{\Delta}^a$ and to
the G transformation operators of form (\ref{35})  with
$U={1\over2}\varepsilon_{ab}[\bar{\Delta}^b,[\bar{\Delta}^a,\Delta]]$,
respectively, the differential expressions for the canonical operators
$\bar{\Delta}^a$ and $\Delta$ coinciding with the corresponding expressions
for the operators $\bar{\nabla}^a$ and $\nabla$. In this case the quantum
action must satisfy, formally, the quantum $Sp(2)$ master equation
$$
\bar{\Delta}^a(\hbar)e^{{i\over\hbar}S}=0,
$$
or
$$
\frac{1}{2}(S,S)^a+\int dx\varepsilon^{ab}\Phi_{Ab}^*(x)
    \frac{\delta}{\delta{\bar\Phi(x)}_{A}}S=i\hbar\Delta_2^aS,
$$
the differential expression for $\Delta_2^a$ coinciding with the differential
expression for $\nabla_2^a$.

{\bf Acknowledgments}

The work of S. S. S. is  supported by Russian Foundation
for Basic Researches under the Grant RFBR--99--02--17916 and by
Human Capital and Nobility Program of the European Community
under the Projects INTAS 96--0308, RFBR-INTAS-95-829.
I. V. T. is partially supported
by Russian Foundation for Basic Researches under the Grant
RFBR--99--01--00980 and by Human Capital and Nobility Program of the
European Community under the Project RFBR--96--0308.

\end{document}